\RequirePackage{fix-cm}
\documentclass[twocolumn,epjc3]{svjour3}  
\smartqed  
\RequirePackage{graphicx}
\usepackage{amsmath}
\usepackage{cases}
\usepackage{amsfonts}
\usepackage{amssymb}
\usepackage{bm}
\usepackage{color}
\usepackage{epsfig}
\usepackage{eucal}
\usepackage{float}
\usepackage{microtype}
\usepackage{subfigure}
\usepackage{tikz}
\usepackage{verbatim}
\usepackage{xcolor}
\usepackage{hyperref}
\usepackage{wasysym}
\usepackage{mathrsfs}

\usepackage{graphicx}
\usepackage{dcolumn}
\usepackage{cuted}
\usepackage{xcolor}

\journalname{Eur. Phys. J. C}
\begin{document}

\title{\boldmath Stochastic gravitational wave background from the collisions of dark matter halos
}


\author{Qiming Yan\thanksref{e1,addr1,addr2,addr4,addr5}
        \and
        Xin Ren\thanksref{e2,addr1,addr2}
        \and
        Yaqi Zhao\thanksref{e3,addr1,addr2}
        \and
        Emmanuel N. Saridakis\thanksref{e4,addr3,addr1,addr2,addr6}
}

\thankstext{e1}{e-mail: yqmtobephd@stu.pku.edu.cn}
\thankstext{e2}{e-mail: rx76@ustc.edu.cn}
\thankstext{e3}{e-mail: zxmyg86400@mail.ustc.edu.cn}
\thankstext{e4}{e-mail: msaridak@noa.gr}


\institute{Deep Space Exploration Laboratory/School of Physical Sciences, University of Science and Technology of China, Hefei, Anhui 230026, China \label{addr1}
           \and
           CAS Key Laboratory for Researches in Galaxies and Cosmology/Department of Astronomy, School of Astronomy and Space Science, University of Science and Technology of China, Hefei, Anhui 230026, China\label{addr2}
           \and
           School of Astronomy and Space Science, University of Chinese Academy of Sciences, Beijing 100049, China \label{addr4}
\and
Kavli Institute for Astronomy and Astrophysics, and School of Physics, Peking University, Beijing, 100871, China \label{addr5}
\and
            National Observatory of Athens, Lofos Nymfon, 11852 Athens, 
Greece\label{addr3}
\and
Departamento de Matem\'{a}ticas, Universidad Cat\'{o}lica del Norte, Avda. Angamos 0610, Casilla 1280 Antofagasta, Chile\label{addr6}
}

\date{Received: date / Accepted: date}

\maketitle

\begin{abstract}
We investigate the effect of the dark matter (DM) halos collisions, namely collisions of galaxies and galaxy clusters, through gravitational bremsstrahlung, on the  stochastic gravitational wave background. We first calculate the gravitational wave signal of a single collision event, assuming point masses and linear perturbation theory. Then we proceed to the  calculation of the energy spectrum of the collective effect of all  dark matter  collisions in the Universe. Concerning the DM halo collision rate, we show that it is given by the product of the number density of DM halos,  which is calculated by the extended Press-Schechter (EPS) theory, with  the collision rate of a single DM halo, which is given by simulation results, with a function of the linear growth rate of matter density through cosmological evolution. Hence, integrating over all mass and distance ranges, we finally extract the spectrum of the stochastic gravitational wave background created by DM halos collisions. As we show, the resulting  contribution to the stochastic gravitational wave background is of the order of $h_{c} \approx 10^{-29}$ in the band of $f \approx 10^{-15} Hz$. However, in very low frequency band, it is larger. With current observational sensitivity it cannot be detected.
\keywords{Stochastic gravitational wave \and gravitational wave source \and pulsar time array}
\end{abstract}

\section{Introduction}

Recently, the gravitational wave (GW) detecting technology has been developing rapidly. In 2015, the detection of binary black holes merger GW150914 by the LIGO experimental cooperation signaled the first detection of gravitational waves \cite{LIGOScientific:2016aoc}, while in 2017, the joint detection of GW170817 \cite{LIGOScientific:2017vwq} and GRB170817A \cite{Goldstein:2017mmi} opened the new era of multi-messenger astronomy \cite{Addazi:2021xuf}. In general, with the increasing amount of detected gravitational wave events \cite{LIGOScientific:2020ibl} one has improved statistics that allows to track the history of the universe \cite{Ezquiaga:2017ekz,Zhao:2017cbb} and impose bounds on various cosmological parameters \cite{LIGOScientific:2017zic,LIGOScientific:2017adf}, as well as constrain various theories of gravity \cite{Baker:2017hug,Farrugia:2018gyz,Cai:2018rzd,Hohmann:2018wxu,CANTATA:2021ktz}. Moreover, for different frequencies and types of gravitational wave sources, various detection means have been designed and implemented. Besides ground-based laser interferometers such as LIGO, Virgo and KAGRA, which probe high frequency bands ($10-10^{4}$ Hz), space-based laser interferometers such as LISA \cite{Amaro-Seoane:2012vvq,LISA:2017pwj,Papanikolaou:2022chm,Domenech:2020ssp} for intermediate frequency gravitational waves ($10^{-4}-1$ Hz), and the pulsar timing array (PTA) \cite{Lentati:2015qwp,NANOGrav:2020bcs,Chen:2022ooe,Chen:2018yje,EPTA:2023sfo,Antoniadis:2023xlr} for lower frequency bands ($10^{-9}-10^{-6}$ Hz), are also raised. These observational avenues allow us to acquire rich information from GWs of different types and sources, among which stochastic gravitational wave background  is attracting increasing interest.

Stochastic gravitational  wave background (GWB) is a type of random background signal that exists in an analogous way to the cosmic microwave background. The contribution of GWB can be roughly divided into cosmological sources and astrophysical sources \cite{Christensen:2018iqi}. Astrophysical originated GWB contains all types of unresolved GW emitting events, including binary black hole mergers \cite{LIGOScientific:2016fpe,Owen:1998xg,Jaffe:2002rt,Stergioulas:2011gd,Clark:2015zxa,DeLillo:2022blw,Agarwal:2022lvk}. These signals can provide information about astrophysical source populations and processes over the history of the universe \cite{Wang:2016tbj,Flauger:2020qyi,Bellomo:2021mer,Zhao:2022cnn}. On the other hand, cosmological originated GWB mainly involves primordial gravitational perturbations during the inflation epoch \cite{Maggiore:1999vm,Ananda:2006af,Caprini:2018mtu}, or perturbations arising from primordial black holes fluctuations \cite{Musco:2012au,Papanikolaou:2021uhe,Papanikolaou:2022hkg,Banerjee:2022xft,Papanikolaou:2020qtd}. GW signals typically remain unaffected during their propagation, and thus they can provide valuable information about the very early stages of the universe. For instance, different inflationary models can lead to different predictions for the GWB spectrum \cite{Dufaux:2007pt,Cai:2007xr,Kuroyanagi:2008ye,Mavromatos:2020kzj,Zhou:2020kkf,Achucarro:2022qrl,Zhu:2021whu,Cai:2022erk,Franciolini:2022tfm,Addazi:2022ukh,Mavromatos:2022yql,Kazempour:2022xzy,Papanikolaou:2022did}, and thus GWB can be used as a probe of this primordial universe epoch. Since GWB can provide us with important astrophysical and cosmological probes, it is crucial to understand its composition and properties \cite{Allen:1997ad,Capozziello:2008fn,Sesana:2008mz,Romano:2016dpx,Huang:2016cjm,Cai:2017aea,Cai:2019jah,Regimbau:2022mdu,NANOGRAV:2018hou,Speri:2022kaq,Georgousi:2022uyt,NANOGrav:2023hfp}.

On the other hand, according to observations, dark matter (DM) constitutes a significant  fraction of the energy density of the universe \cite{Planck:2018nkj,Planck:2018vyg,Planck:2013pxb}. Its microphysical nature and possible interactions remain unknown \cite{Bertone:2004pz,Jungman:1995df,Preskill:1982cy,Arkani-Hamed:2008hhe}, nevertheless we do know unambiguously that DM takes part in gravitational  interaction \cite{Rubin:1980zd,Clowe:2006eq}. Current theory predicts that the main part of DM is concentrated in dark halos, which coincide in position with galaxy or galaxy clusters \cite{Navarro:1995iw}. These galaxies and galaxy clusters, and thus dark halos too, are typically accelerating and merging through their mutual attraction \cite{Davis:1985rj,White:1977jf,Bullock:1999he}. Such processes  can in principle release GW signal through gravitational bremsstrahlung \cite{Peters:1970mx,Crowley:1977us,Smarr:1977fy,Kovacs:1978eu,Farris:2011vx,Mougiakakos:2021ckm,Herrmann:2021lqe,Mougiakakos:2022sic,Jakobsen:2021lvp,Gondan:2021fpr,Riva:2022fru,Inagaki:2010nu,Inagaki:2012th}.  This process can be approximately described as an elastic collision between two particles. The approximate calculation results of the gravitational waves released during the collision process have been obtained previously in some literature \cite{2010gfe..book.....M}. However, it is important to note that simply considering the gravitational waves released during the elastic collision process is the ideal hypothesis because the release of gravitational waves takes away the system's energy, making the collision no longer elastic. Therefore, the gravitational wave spectrum calculated in this way can only be applicable below a certain cutoff frequency.

In this work, we are interested in investigating for the first time the possible GW signals that could be emitted through bremsstrahlung during dark halo merger and collisions, and their contribution to the stochastic GWB. In particular, we will first consider a single event of two DM halos collision, and we will calculate the emitted GW signal. Then, we will  calculate the energy spectrum contribution to the stochastic GWB, taking the DM halo collision rate into consideration. The structure of the article is as follows. In Section \ref{sec: single case} we analyze the GW emitted during the collision of two galaxies or two galaxy clusters. In Section \ref{sec: SGWB} we integrate over redshift and DM halos parameters to extract the contribution to stochastic GWB. Finally, in Section \ref{sec: conclusion}  we conclude and discuss our results.

\section{Gravitational waves  emitted during  a single collision}
\label{sec: single case}

In this section, we aim  at estimating the gravitational waves emitted during a single collision event. In particular, we calculate the GW radiated by the collision of two DM halos, which corresponds to the collision of two galaxies or two galaxy clusters. 

\cite{2010gfe..book.....M} considered the gravitational waves released during the elastic collision between two particles. In section 4.4.1 of \cite{2010gfe..book.....M}, the energy-momentum tensor was approximated directly using the 4-momentum of the particles before and after the collision without considering collision process.

In the case where the collision velocity is much lower than the speed of light $c$, let A and B be the two particles participating in the collision, $m_{Q}$, $p_{Q}^{\mu}, {p_{'} }_{Q}^{\mu} $ and $\vec{v}_{Q},\vec{v}_{Q}^{'} $ be the mass, 4-momenta and velocities of particle Q before and after the collision. Then, before the collision occurs, the energy-momentum tensor of the two particles is given by $ \sum_{Q=A,B} 
\frac{p_{Q}^{\mu} p_{Q}^{\nu} }{ m_{Q}   } \delta^{(3)} ( \vec{x}_{Q} - \vec{v}_{Q} t  ) $. 
Treating the collision process as instantaneous, the collision occurs at $t = 0$. 
Afterwards, the energy-momentum tensor of the two particles is given by $ \sum_{Q=A,B}  \frac{ {p^{'} }_{Q}^{ \mu} {p^{'} }_{Q}^{ \nu} }{ m_{Q}   } \delta^{(3)} ( \vec{x}_{Q} - \vec{v}^{'}_{Q} t  ) $. Let $\theta(t)$ be the unit step function. In this way, the energy-momentum tensor of the particles throughout the entire process can be written as
\begin{align}
T^{\mu \nu} \approx \sum_{Q=A,B}  
\frac{p_{Q}^{\mu} p_{Q}^{\nu} }{ m_{Q}   } \delta^{(3)} ( \vec{x}_{Q} - \vec{v}_{Q} t  ) \theta(-t) + \notag \\   \frac{ {p^{'} }_{Q}^{ \mu} {p^{'} }_{Q}^{ \nu} }{ m_{Q}   } \delta^{(3)} ( \vec{x}_{Q} - \vec{v}^{'}_{Q} t  ) \theta(t) ~.
\end{align}
Based on this energy-momentum tensor, the gravitational wave  energy spectrum can be further calculated. In the next section, we will see the energy spectrum given in this article is consistent with that in \cite{2010gfe..book.....M} at relatively high frequencies. However, since we give the precise calculation of the particle motion during the entire collision process, in our calculation we also give the contribution of the lower-frequency gravitational waves compared to the results given in \cite{2010gfe..book.....M}.

Besides, at very high frequencies, the physical processes corresponding to the emission of high-frequency gravitational waves are not reflected in either analytical method and cannot be dealt with. As a result, the gravitational wave spectra obtained from both methods can only be applicable below a certain cut-off frequency. In fact, without using numerical relativity for high precision calculations, any gravitational wave spectra obtained through analytical methods require an artificially estimated cut-off frequency. In the next section, we will see that the spectra obtained from both methods diverge when integrated to arbitrarily high frequencies, thus necessitating the establishment of a cut-off frequency.

The following is a more detailed and accurate calculation process we carried out.
According to observations,  such a collision typically has a huge duration, which in turn implies that the energy radiated through GWs per unit time is not very large, and thus we can safely use linear perturbation theory in the involved calculations. Specifically, we use 
\cite{Maggiore:2007ulw}
\begin{align}
&
    g_{\mu \nu} = \eta_{\mu \nu}+h_{\mu \nu}, \quad\left|h_{\mu \nu}\right| \ll 1, \\
&
    \bar{h}_{ij}(t, \mathbf{x})=\frac{2 G}{r c^4} \frac{\mathrm{d}^{2} I_{i j}\left(t_{r}\right)}{\mathrm{d} t^{2}}, \quad t_{r}=t- \frac{r}{c}, \label{h}
\end{align}
where $G$ is the gravitational constant, $c$ is the speed of light,  and $r$ is the distance from us to the center of mass of the two galaxies or galaxy clusters. Moreover,  $I_{i j}$ is the quadruple moment
\begin{equation}
    I_{i j}(t)=\int y^{i} y^{j} T^{00}(t, \mathbf{y}) d^{3}  y=\int y^{i} y^{j} \rho(t, \mathbf{y}) d^{3} y, \quad 
\end{equation}
where $T^{\mu \nu}$ is energy-momentum tensor, $\rho$ is energy density, and $y^{i}$ is the spatial coordinate. Since the goal of our calculation is to acquire an estimation of the order of the magnitude of the resulting signal, we can consider these two DM halos as mass points, with mass $M_a$ and position $\mathbf{y}_{(a)}(t)$ at   time $t$. Hence, the density $\rho$ can be written as 
\begin{align}
    \rho(t, \mathbf{y})= \sum_{a} M_a \delta^{3}(\mathbf{y} - \mathbf{y_{(a)}}(t) ),
\end{align}
while the quadruple moment $I_{i j}(t)$ becomes 
\begin{align}
    I_{i j}(t)= \int y^{i} y^{j} \rho(t, \mathbf{y}) d^{3} y = \sum_{a} M_a  y^{i}_{(a)}(t) y^{j}_{(a)}(t).
\end{align}
Finally, since  the relative speed of two galaxies or galaxy clusters is much smaller than the speed of light, we can use Newtonian mechanics to handle their dynamics.

\begin{figure}[ht]
    \centering
    \includegraphics[width=1\columnwidth]{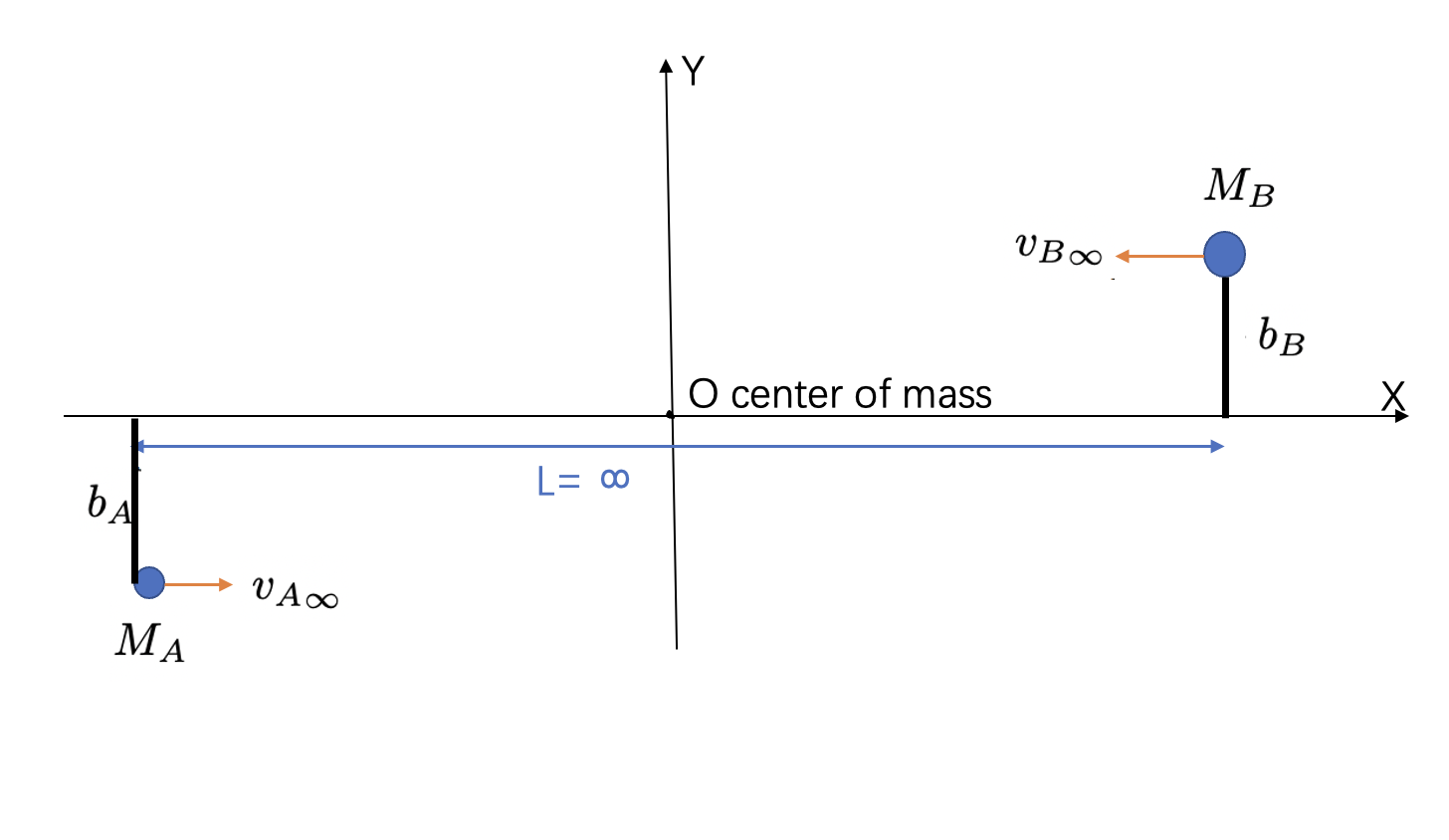}
    \caption{\it{ An illustrative representation of the initial conditions of the collision. The two galaxies or galaxy clusters are considered as points with masses  $M_{A}$ and $M_{B}$, where $ b_{A}$ and $b_{B}$ are the impact parameters.}}
    \label{fig:collisions}
\end{figure}

For simplicity, we  write the equations in the center-of-mass frame of these two mass points. By definition, we have  
\begin{align}
    M_{A} \mathbf{r}_{A} + M_{B} \mathbf{r}_{B} = \mathbf{0}, \label{center}
\end{align}
where $M_{A}$, $M_{B}$ are the masses of the mass points A and B,  with $\mathbf{r_{A}}$, $\mathbf{r_{B}}$ their position vectors. From Newtonian mechanics we have
\begin{align}
    \ddot{\mathbf{r}}_{A} = - \frac{G M_{B}}{|\mathbf{r}_{A} - \mathbf{r}_{B} |^{2}} \frac{\mathbf{r}_{A}}{|\mathbf{r}_{A}|}, 
\end{align}
which using \eqref{center} gives
\begin{align}
    \ddot{\mathbf{r}}_{A} &=   -\mu_{B}  \frac{\mathbf{r}_{A}}{|\mathbf{r}_{A}|^3},
\end{align}
where we have defined $\mu_{B} \equiv \frac{GM_{B}}{(1 + \frac{M_{A}}{M_{B}})^2} $. Additionally, we assume that the two points are initially at infinite distance, their relative speed is $v_{\infty} = {v_{A}}_{\infty} +  {v_{B}}_{\infty}$, and the impact  parameter is $b = b_{A} + b_{B}$. From Newtonian mechanics we know that the trajectory of each point is a hyperbola and the two points are moving in a plane (we set this plane as $z=0$ plane, and thus $\mathbf{r}_{A} = (x_{A},y_{A},0)$), while the total energy of the system  is  positive. Additionally,  the mass center of  these two DM halos will not follow a hyperbolic trajectory at all  times, in order to acquire a collision. In Fig. \ref{fig:collisions} we depict an illustrative representation of the initial conditions of the collision.

Let us start with the   beginning of the collision, when the two DM halos start moving towards  each other. For point A we have
\begin{align}
    (x_{A}+a_{A}e_{A})^2 - (y_{A})^2 = a^{2}_{A},
\end{align}
where 
\begin{gather}
    a_{A} = \frac{\mu_{B}}{({v_{A}}_{\infty})^2}, \\
    e_{A}  =\sqrt{1 + \frac{{v_{A}}^4_{\infty} 
    b^{2}_{A}}{(\mu_{B})^2}}, \\
    a = a_A + a_B.
\end{gather}
We proceed by defining  $\lambda_{A}$ through
\begin{align}
    e_{A}  \sinh(\lambda_{A}) - \lambda_{A} = \frac{{v_{A}}_{\infty} t }{a_{A}},
\end{align}
hence 
\begin{align}
    \label{xA1}
    r^{1}_{A}=x_{A} &= a_{A}  \left[e_A - \cosh(\lambda_{A})\right],\\
    r^{2}_{A}=y_{A} &= a_{A}  \left[\sqrt{e^{2}_{A} - 1} \sinh(\lambda_{A})\right].
    \label{yA1}
\end{align}
Note that $t=0$ corresponds to the time when the two mass points have the shortest distance.

In order to obtain the GW amplitude $h_{ij}$, we proceed to the  calculation of  the quadrupole moment $I_{i j}(t)$ and its second time derivative. We have
\begin{eqnarray}
    I_{ij} &&= M_{A} r^{i}_{A} r^{j}_{A}  + M_{B} r^{i}_{B} r^{j}_{B},
    \\
    \frac{\mathrm{d}^{2} I_{i j}}{d t^2} &&= M_{A} (\ddot{r}^{i}_{A} r^{j}_{A} + r^{i}_{A} \ddot{r}^{j}_{A} + 2 \dot{r}^{i}_{A} \dot{r}^{j}_{A} )\notag
    \\
    &&
    + M_{B}(\ddot{r}^{i}_{B} r^{j}_{B} + r^{i}_{B} \ddot{r}^{j}_{B} + 2 \dot{r}^{i}_{B} \dot{r}^{j}_{B} ).\label{dIij1}
\end{eqnarray}
From \eqref{xA1}, \eqref{yA1} we find
\begin{align}
    \dot{x}_{A} &= -\frac{a_{A} \sinh (\lambda_{A})}{\sqrt{\frac{a^3_{A}}{\mu_{B}}} [e_{A} \cosh (\lambda_{A})-1]},\\
    \ddot{x}_{A} &= \frac{\mu_{B} (\cosh (\lambda_{A})-e)}{a^2_{A} [e_{A} \cosh (\lambda_{A})-1]^3},\\
    \dot{y}_{A} &= \frac{a_{A} \sqrt{e^2_{A}-1} \cosh (\lambda_{A})}{\sqrt{\frac{a^3_{A}}{\mu_{B}}} [e_{A} \cosh (\lambda_{A})-1]},\\
    \ddot{y}_{A} &= -\frac{\sqrt{e^2_{A}-1} \mu_{B} \sinh (\lambda_{A})}{a^2_{A} [e_{A} \cosh (\lambda_{A})-1]^3},
\end{align}
and thus inserting into \eqref{dIij1} we extract all the second time derivatives of the quadrupole moment $I_{i j}(t)$, namely
\begin{small}
\begin{eqnarray}
    \frac{\mathrm{d}^{2} I_{11}}{d t^2}&&
    =  
    \frac{\mu_{B} M_{A} }{2 a_{A} [e_{A} \cosh (\lambda_{A})-1]^3} 
     \\
    && \!\!\!\!\!\!\!\!\!\!\!\!\!\!
    \times\{7 e_{A} \cosh (\lambda_{A})+e_{A} [\cosh (3 \lambda_{A})-4 e_{A}]-4 
    \cosh (2 \lambda_{A})\} 
    \notag \\
    && \!\!\!\!\!\!\!\!\!\!\!\!\!\!
    + \frac{\mu_{A} M_{B} }{2 a_{B} [e_{B} \cosh (\lambda_{B})-1]^3} 
    \notag \\
    && \!\!\!\!\!\!\!\!\!\!\!\!\!\!
    \times\{7 e_{B} \cosh (\lambda_{B})+e_{B} [\cosh (3 \lambda_{B})-4 
    e_{B}]-4 \cosh (2 \lambda_{B})\},
    \notag \\
    \frac{\mathrm{d}^{2} I_{12}}{d t^2} &&
    = 
    -\frac{\sqrt{e^2_{A}-1} \mu_{B} M_{A} \sinh (\lambda_{A}) }{a_{A} [e_{A} \cosh (\lambda_{A})-1]^3} 
    \\
    && \!\!\!\!\!\!\!\!\!\!\!\!\!\!
    \times\{e_{A} [\cosh (2 \lambda_{A})+3]-4 \cosh (\lambda_{A})\}
    \notag \\
    && \!\!\!\!\!\!\!\!\!\!\!\!\!\!
    -\frac{\sqrt{e^2_{B}-1} \mu_{B} M_{A} \sinh (\lambda_{B}) }{a_{B} [e_{B} \cosh (\lambda_{B})-1]^3}  \notag\\
    && \!\!\!\!\!\!\!\!\!\!\!\!\!\!
    \times \{e_{B} [\cosh (2 
    \lambda_{B})+3]-4 \cosh (\lambda_{B})\},
    \notag \\
    \frac{\mathrm{d}^{2} I_{22}}{d t^2} 
    &&
    = 
    \frac{\left(e^2_{A}-1\right) \mu_{B} M_{A} }{2 a_{A} [e_{A} \cosh (\lambda_{A})-1]^3}
    \\
    && \!\!\!\!\!\!\!\!\!\!\!\!\!\!
    \times[3 e_{A} \cosh (\lambda_{A})+e_{A} \cosh (3 \lambda_{A})-4 \cosh (2 \lambda_{A})]
    \notag\\
    && \!\!\!\!\!\!\!\!\!\!\!\!\!\!
    + \frac{\left(e^2_{B}-1\right) \mu_{A} M_{B} }{2 a_{B} [e_{B} \cosh 
    (\lambda_{B})-1]^3}
    \notag\\
    && \!\!\!\!\!\!\!\!\!\!\!\!\!\!
    \times[3 e_{B} \cosh 
    (\lambda_{B})+e_{B} \cosh (3 \lambda_{B})-4 \cosh (2 \lambda_{B})].\notag
\end{eqnarray}
\end{small}

For simplicity, we define $M \equiv M_{A} + M_{B}$, and the mass ratio $x \equiv M_{A}/M_{B}$. Noting that we have $M_{A} b_{A} = M_{B} b_{B}$ and $M_{A} v_{A \infty} = M_{B} v_{B \infty}$, so that $1/x = b_{A}/b_{B} = v_{A \infty}/ v_{B \infty}$.

As a result, we get 
$M_{A} = \frac{M}{(1+1/x)}, M_{B} = \frac{M}{1+x} ,b_{A} =  \frac{b}{1+x}, b_{B}=\frac{b}{1+1/x},v_{A \infty} = \frac{v_{\infty}}{1+x}, v_{B \infty} = \frac{v_{\infty}}{1+1/x}$.
Substitute the above equations into the formula of 
$\mu_{B},a_{A},e_{A}$,we have
\begin{align}
\mu_{B} = \frac{GM}{(1+x)^3} \\
a_{A} = \frac{GM}{v^{2}_{\infty}} \frac{1}{1+x} \\
e_{A} = \sqrt{1+ (\frac{b v^2_{\infty}}{GM})^2} 
\end{align}
We note that $e_{A}$ is independent from mass ratio $x$, so $e_{A} = e_{B}$. Besides, $v_{A \infty} /a_{A} = v^{3}_{\infty} /GM $ is also  independent from mass ratio $x$, so the $\lambda_{A}$  is  just equal to $\lambda_{B}$.In short, the mass ratio $x$  only affects $\ddot{I}_{ij}$ by factors like $\mu_{B}M_{A}/a_{A} = Mv^{2}_{\infty} x/(1+x)^3$.Therefore, we can get the results in any mass ratio from the equal mass result by the following equation,
\begin{align}
\ddot{I}_{ij}(M,x) = 4[\frac{x}{(1+x)^3} + \frac{1/x}{(1+1/x)^3} ]\ddot{I}_{ij}(M,x=1)
\end{align}

\begin{figure*}[ht]
    \centering
    \includegraphics[width=0.64\columnwidth]{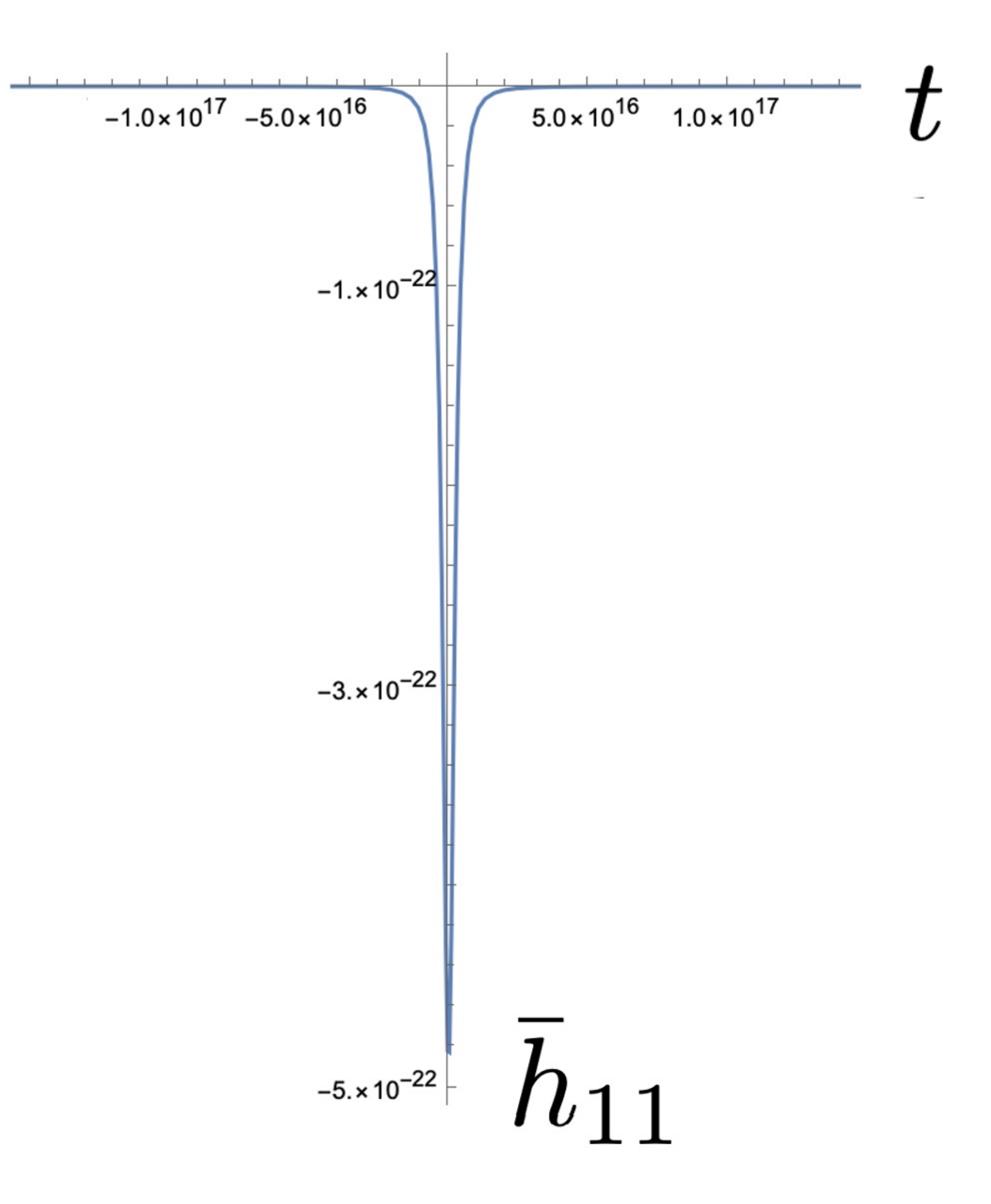}
    \includegraphics[width=0.64\columnwidth]{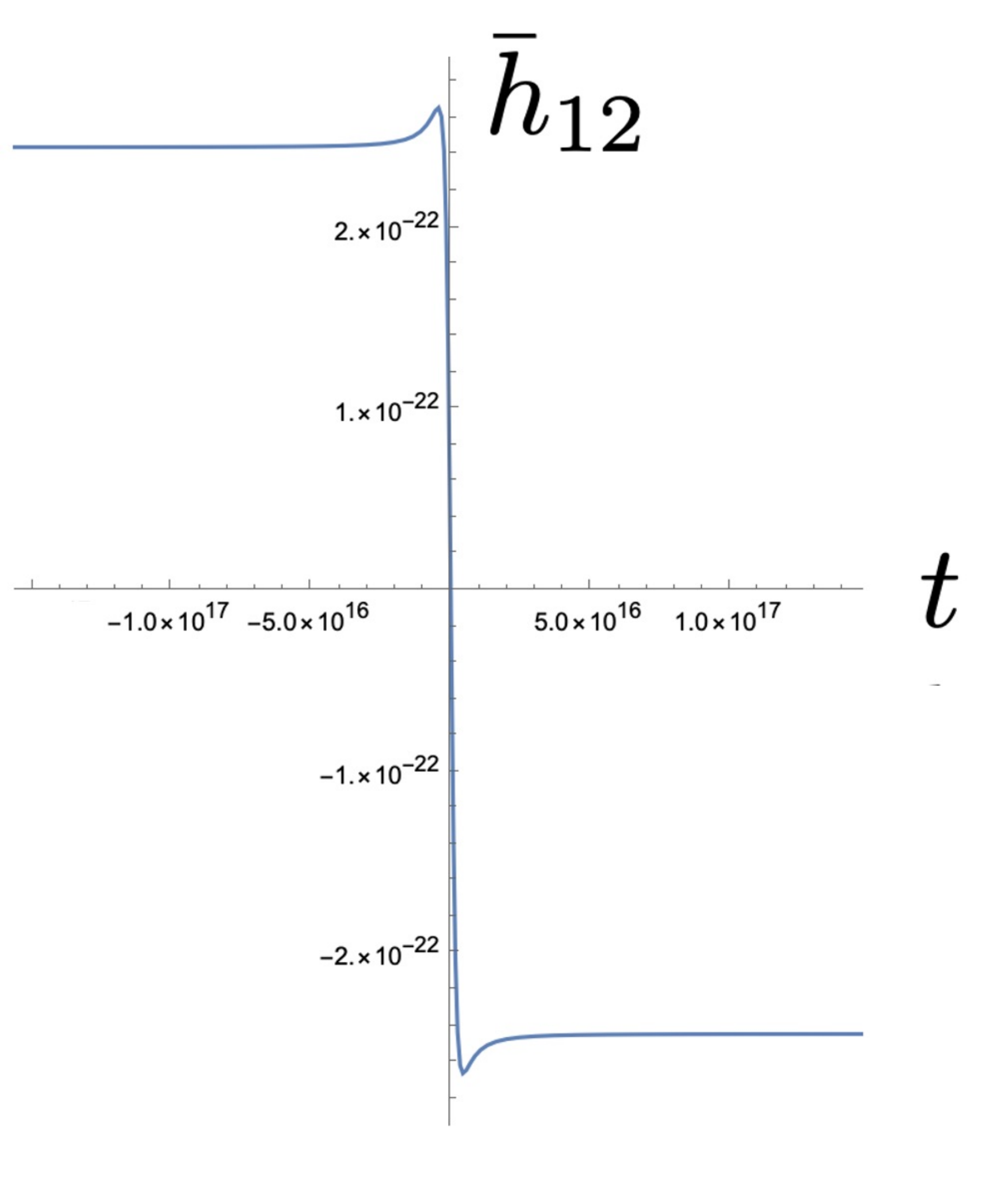}
    \includegraphics[width=0.64\columnwidth]{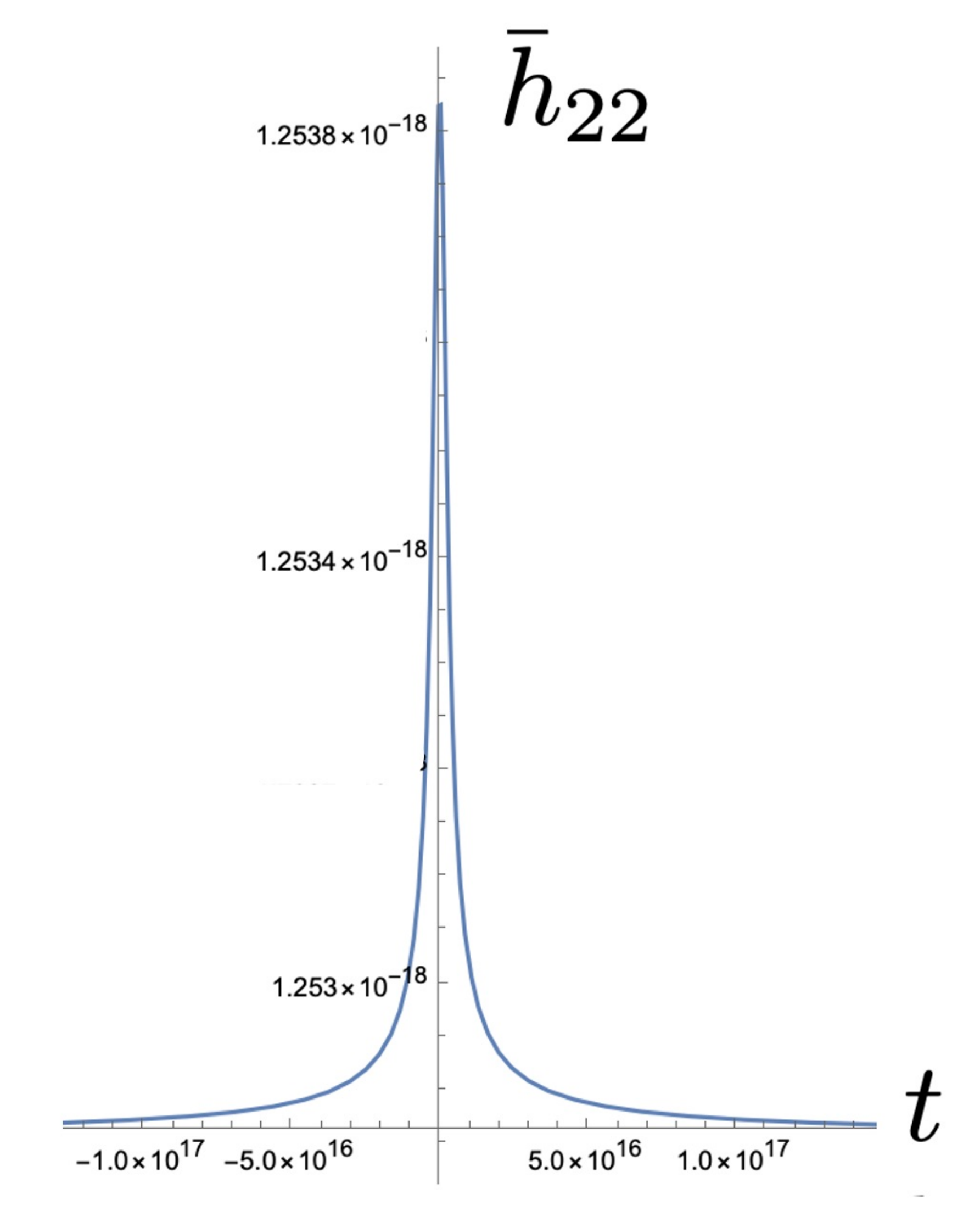}
    \caption{\it{
    The dimensionless components of the gravitational wave signal arising from a single event of the collision of two DM halos, i.e., the collision of two galaxies or clusters of galaxies. The left panel shows the $\bar{h}_{11}$ component, the middle panel the $\bar{h}_{12}$ component and the right panel the $\bar{h}_{22}$ component. The time $t=0$ corresponds to the shortest distance between  the two DM halos, that is   the moment in which  $\bar{h}_{11}$   and $\bar{h}_{22}$   reach their peaks  and $\bar{h}_{12}$  exhibits the largest variation. We have imposed the typical values $M_{A} = M_{B} = 10^{9} M_{\odot}$,  ${v_{A}}_{\infty} = {v_{B}}_{\infty} = 300 km/s$, $b_{A} = b_{B} = 10^{4} ly$, and we have assumed that the distance from Earth is       $\sim10^9 ly$. Time $t$ is measured in  seconds.}}
    \label{fig: GW component}
\end{figure*}

We can now use \eqref{h} in order to obtain  the GW signal in the time domain. As typical values we set $M_{A} = M_{B} = 10^{9} M_{\odot}$, namely the order of  mass of a (dwarf) galaxy, where $M_{\odot}$ is the mass of the Sun, and we use ${v_{A}}_{\infty} = {v_{B}}_{\infty} = 300 km/s$, $b_{A} = b_{B} = 10^{4} ly$, which are the typical values for   galaxy collisions. Moreover, we assume that the collision happens at a distance of $10^9 ly$ from the Earth, which is roughly the  distance  of the source of GW150914. Hence, we can   estimate the magnitude of the GW signal. In Fig. \ref{fig: GW component} we present the obtained dimensionless GW signal $\bar{h}_{ij}$,  as a function of time $t$. Since $t=0$ corresponds to the time of shortest distance, the change rate of $\bar{h}_{ij}$ is fastest at this time, as expected. As we observe, the variation of $\bar{h}_{ij}$ is of the order of $5 \times 10^{-22}$ during the collision. However, this variation corresponds to a large time scale (about $10^{15} s$), which implies that a single signal of this kind of GW is extremely hard to be detected. Additionally, we can see that the evolution of $\bar{h}_{12}$ is faster than that of $\bar{h}_{11}$, $\bar{h}_{22}$, which implies that $\bar{h}_{12}$ will be dominant in relatively higher frequency than that of $\bar{h}_{11}$, $\bar{h}_{22}$.

The GW amplitude
in TT gauge $h^{TT}$ is the traceless version of $\bar{h}_{ij}$.We can project the GW amplitude
in TT gauge by 
\begin{align}
 h^{TT}_{11} = \frac23 \bar{h}_{11} - \frac13 \bar{h}_{22} ,\\
    h^{TT}_{22} = - \frac13 \bar{h}_{11} + \frac23 \bar{h}_{22} ,\\
    h^{TT}_{33} = - \frac13 \bar{h}_{11} - \frac13 \bar{h}_{22} ,\\
    h^{TT}_{21}=h^{TT}_{12} = \bar{h}_{12},
\end{align}
while all other $h^{TT}_{i j} $ are  equal to zero.

\begin{figure*}[!ht]
    \centering
    \includegraphics[width=0.8\columnwidth]{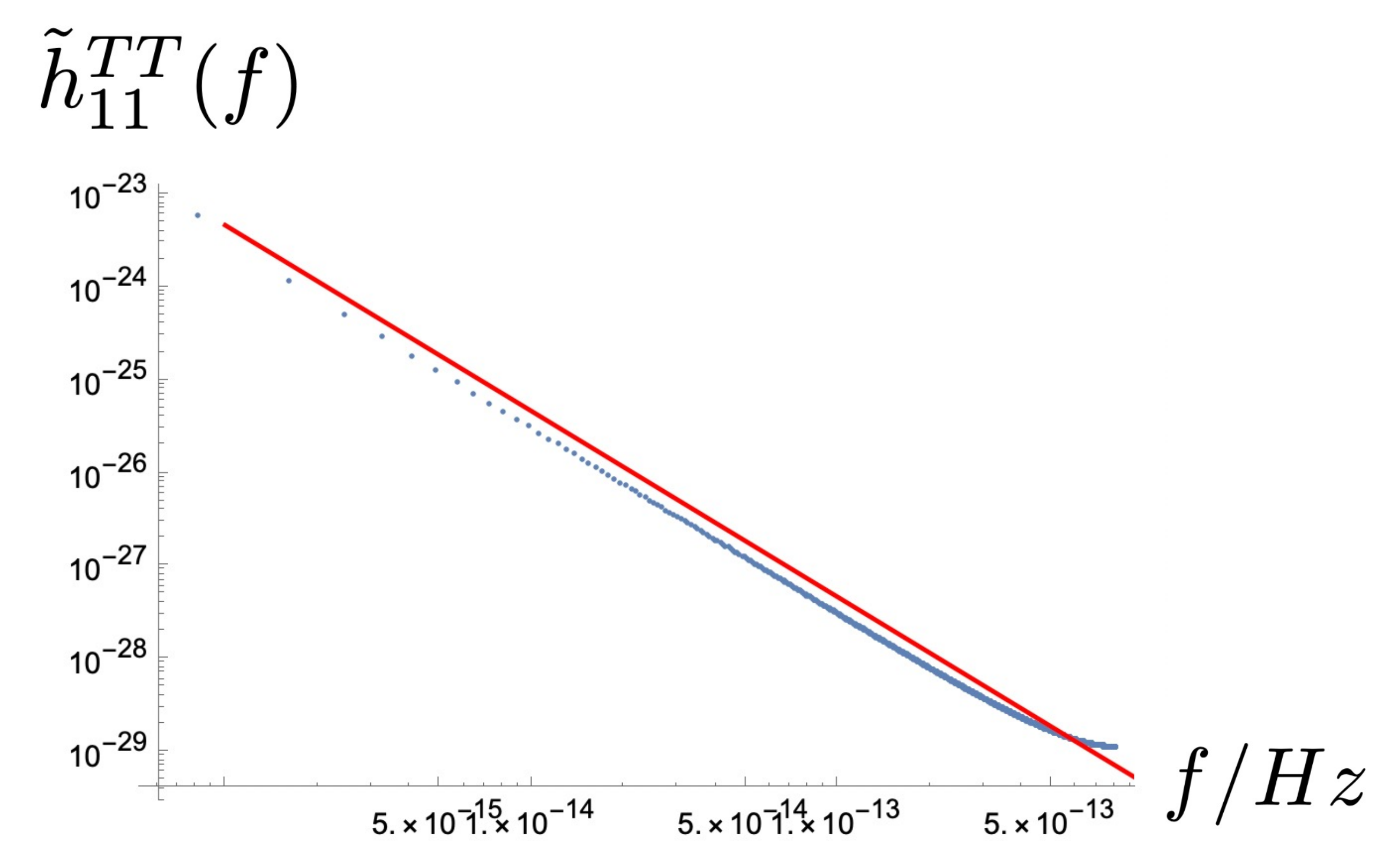}
    \includegraphics[width=0.8\columnwidth]{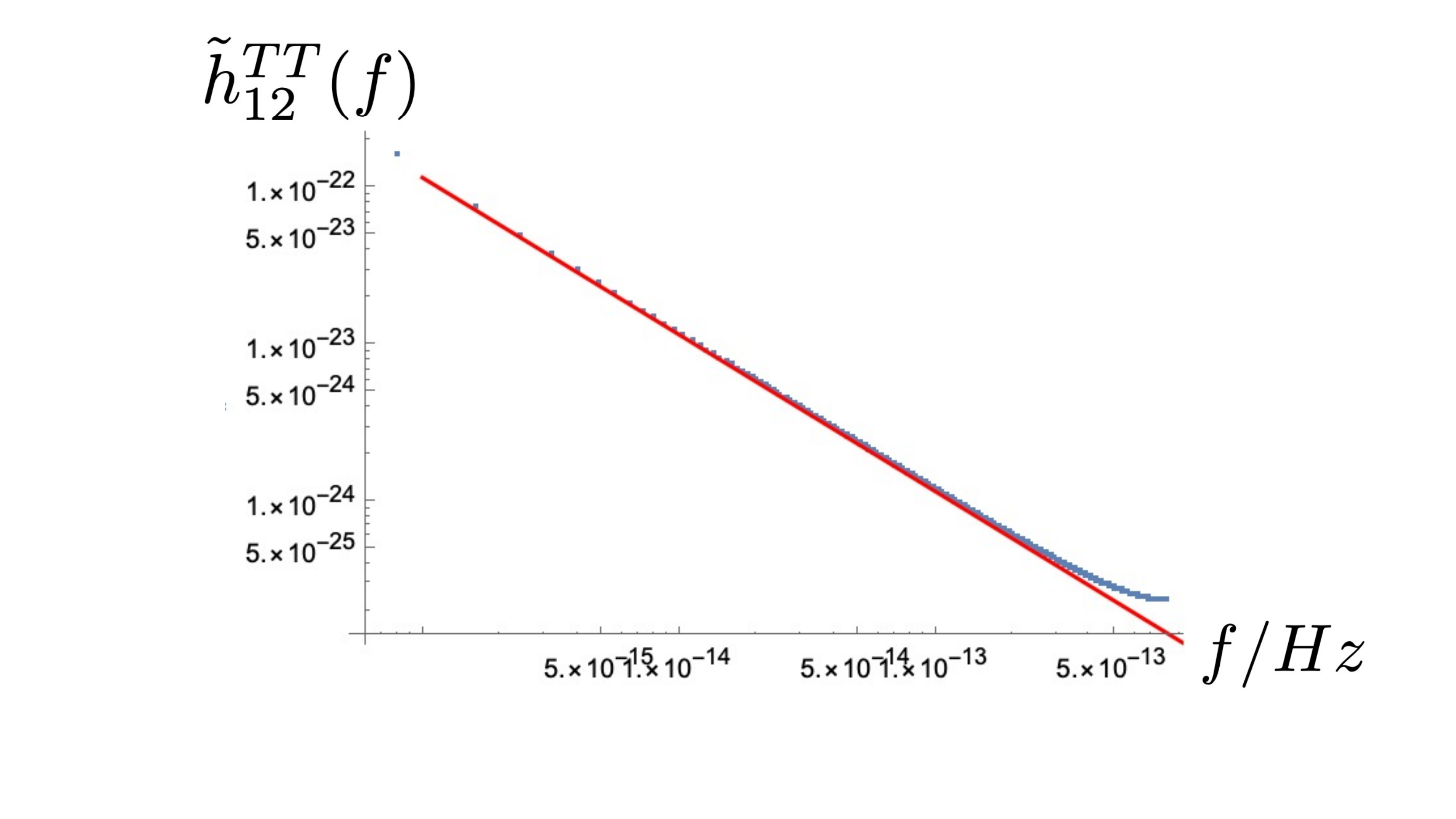}
    \includegraphics[width=0.8\columnwidth]{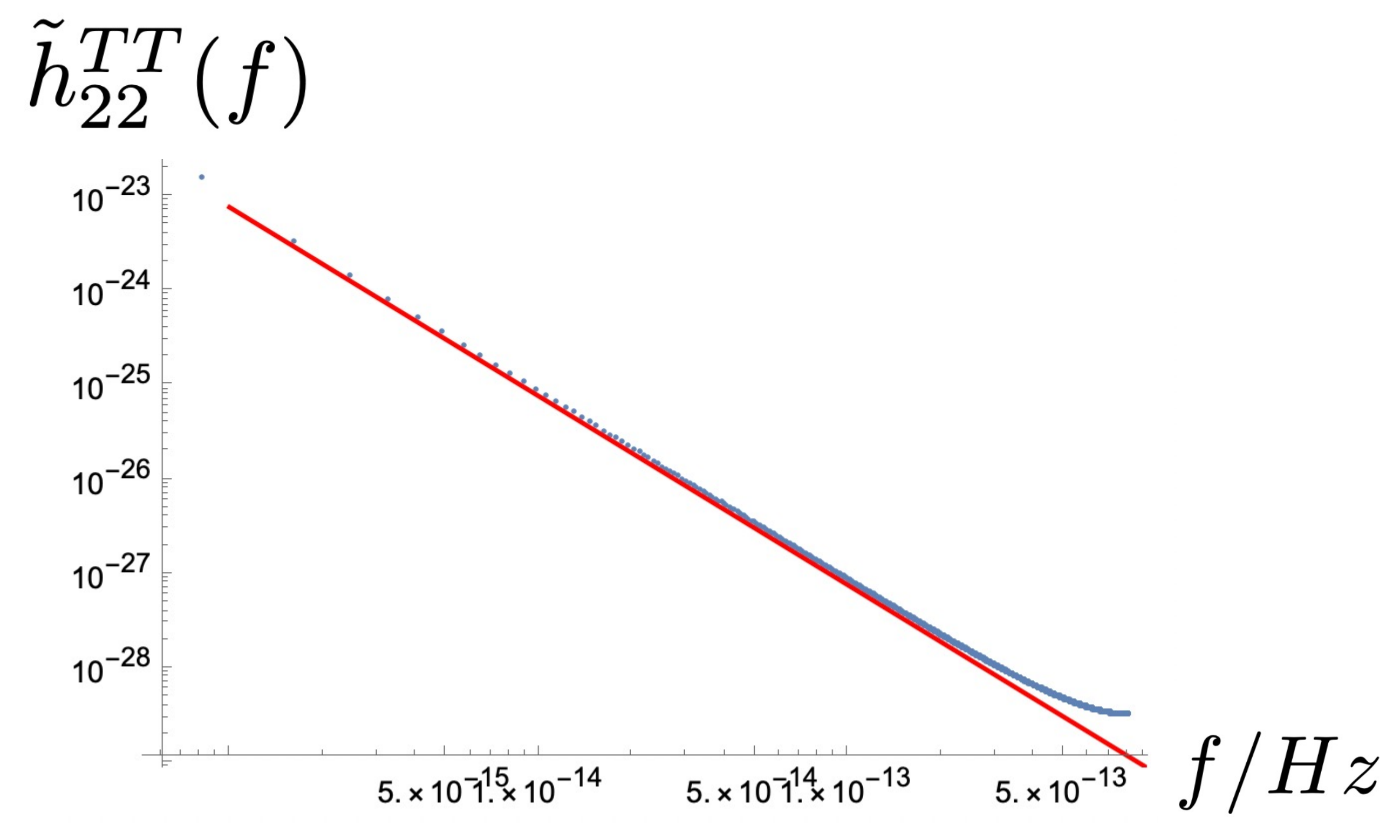}
    \caption{\it{The spectrum of the gravitational waves as a function of the frequency. The upper left panel shows the $\tilde{h}^{TT}_{11}(f)$ component, the upper right panel the $\tilde{h}^{TT}_{12}(f)$ component and the right panel the $\tilde{h}^{TT}_{22}(f)$ component. The blue dots represent the exact results at the time of shortest distance, while the red solid curves are power-law fits, specifically $\tilde{h}^{TT}_{11}(f)\approx 4.7 \times 10^{-54} (Hz/f)^2 $, $\tilde{h}^{TT}_{12}(f)\approx 1.2 \times 10^{-38} (Hz/f)^2$,  $\tilde{h}^{TT}_{22}(f)\approx 7.7 \times 10^{-54} (Hz/f)^2 $.}}
    \label{fig: GW spectrum}
\end{figure*}

We proceed by taking the  Fourier transformation of $\bar{h}_{ij},h^{TT}_{ij}$, in order to investigate its spectrum. In particular, we use
\begin{equation}
    \tilde{\bar{h}}_{ij}(\omega)= \int^{t=+\infty}_{t=-\infty} \mathrm{d} t\, e^{i \omega t}\, \bar{h}_{ij}(t)~,
\end{equation}
\begin{equation}
    \tilde{h}^{TT}_{ij}(\omega) = \int^{t=+\infty}_{t=-\infty} \mathrm{d} t\, e^{i \omega t}\, h^{TT}_{ij}(t)
\end{equation}
where $\omega = 2 \pi f$, with $f$ the frequency. $\tilde{\bar{h}}_{ij}(f)$ obey the power law in a very good approximation for a very wide frequency range. Besides, as $\tilde{\bar{h}}_{11},\tilde{\bar{h}}_{22} \propto  1/f^2$, while $\tilde{\bar{h}}_{12} \propto 1/f$, we can infer that $\tilde{\bar{h}}_{11},\tilde{\bar{h}}_{22}$ will be dominant in the low frequency band while $\tilde{\bar{h}}_{12}$ will be dominant in relatively high frequencies.    In   Fig. \ref{fig: GW spectrum} we present the   dependence of $\tilde{h}^{TT}_{ij}(\omega)(f)$ on  $f$. 

\section{Effect on the stochastic gravitational wave background}
\label{sec: SGWB}

In this section, we calculate  the contribution of the DM halos collisions to the stochastic gravitational wave background. Specifically, we integrate the gravitational wave spectrum of a single collision event over the number density of GW sources.

In principle, in order to compare a theoretical model with observations, one uses both the fractional energy density spectrum $\Omega_{gw}(f)$, as well as the  characteristic strain amplitude $h_{c}(f)$ \cite{Romano:2016dpx}. They are related to the energy spectrum of GWB through the expression
\begin{align}
    \frac{\pi c^2}{4 G} f^{2} h_{c}^{2}(f)
    = \rho_c \Omega_{gw}(f)
    =\frac{d \rho_{\mathrm{gw}}(f)}{d \ln f}.
\end{align}
where $f$ is the frequency of GW    detected on Earth, and $\rho_{c} \equiv 3 c^{2} H_{0}^{2} / 8 \pi G $ is the critical energy density. The  energy spectrum of the stochastic GWB, $\frac{d \rho_{\mathrm{gw}}}{d \ln f}$, can be written as
\begin{align}
    \frac{d \rho_{\mathrm{gw}}(f)}{d \ln f}=\left.\int_{0}^{\infty} d z  \frac{1}{1+z}  \int d\xi \frac{d n}{d z d \xi} \frac{d E(\xi)_{\mathrm{gw}}}{d \ln f_{r}}\right|_{f_{r}=f(1+z)},
\end{align}
with $z$ the redshift at the GW emission. Additionally,  $\frac{d E(\xi)_{gw}}{d \ln(f_r)}$ is  the energy spectrum of a single GW event, which is calculated through the analysis of subsection \ref{sec: single case}, and $f_r$ is the GW frequency in the rest frame of GW sources, and thus $f_r = (1+z) f$. 

We mention that we denote the parameters related to the number density of GW sources collectively by  $\xi=\left\{\xi_{1}, \ldots, \xi_{m}\right\}$, and therefore $\frac{d n}{d \xi_{1} \ldots d \xi_{m} dz} d \xi_{1} \ldots d \xi_{m} d z \equiv \frac{d n}{d \xi dz} d \xi d z$ is the number  density of sources in the redshift interval $[z, z+d z]$ and with source parameters in the interval  $[\xi, \xi+d \xi] $. Hence, in the simple single event of two DM halos collision of the previous section we have $\xi = \{ M, x, v_{\infty} , b \}$, where $M = M_A + M_B$, $x = M_A/M_B$, $v_{\infty} = {v_{A}}_{\infty} +  {v_{B}}_{\infty}$   and  $b = b_{A} + b_{B}$.
 
Let us now calculate the full distribution function $\frac{d n}{d z d \xi} = \frac{d n}{d z dM dx d v_{\infty} d b}$. As we have checked numerically,  the variance of $b,v_{\infty}$ has a minor effect on the final result, not affecting the order of magnitude. Hence, it is a good approximation to  omit the change of $b,v_{\infty}$, and  consider that $\xi = \{M,x\}$. Hence, we have

\begin{align}
    \!\!\!\!\frac{d \rho_{\mathrm{gw}}(f)}{d \ln f}&=\left.\int_{0}^{\infty} d z  
    \frac{1}{1+z}  \int d\xi 
     \frac{d n}{d z d\xi} \frac{d E(\xi)_{\mathrm{gw}}}{d \ln 
    f_{r}}\right|_{f_{r}=f(1+z)}  
    \notag\\
    & \approx \int_{0}^{ 10} d z  \frac{1}{1+z} 
      \int_{M_{min} = 10^9 M_{\odot}}^{M_{max} = 10^{15} M_{\odot}} dM 
      \int_{x_{min}=1}^{x_{max} = 10^5} dx
    \notag\\
    &
     \times\left.\frac{d n}{d z dM dx} \frac{d E(\xi)_{\mathrm{gw}}}{d \ln f_{r}}\right|_{f_{r}=f(1+z)},
    \label{finaleqshort}
\end{align}
where the varying range of $M$ and $x$ is taken from  \cite{Genel:2010pb}.

In the following subsections we will separately calculate the energy spectrum of a single GW event $\frac{d E(\xi)_{\mathrm{gw}}}{d \ln f_{r}}$, and the number density of GW sources $\frac{d n}{d z dM dx}$.

\subsection{Energy spectrum of a single GW event} \label{subsec:energy}

The energy density of a single GW event can be calculated from the (traceless) second  time derivative of the quadrupole moment, namely \cite{Maggiore:2018sht}
\begin{equation}
    \frac{d E(\xi)_{\mathrm{gw}}}{d \ln f_r}
    \approx 
    f_r \frac{2G}{5c^5} (2 \pi f_r)^2 
    (\ddot{\tilde{Q}}_{i  j} (M,x;f_r)) (\ddot{\tilde{Q}}_{i  j} (M,x;f_r)) ,
    \label{singleenergy}
\end{equation}
where  $Q_{i  j}$ is the traceless quadrupole moment  and $\ddot{\tilde{Q}}_{i  j}$ is the  Fourier transformation of the  second time derivative of $Q_{i  j}$, which is related to $I_{ij}$ via 
\begin{gather}
    Q_{11} = \frac23 I_{11} - \frac13 I_{22} ,\\
    Q_{22} = - \frac13 I_{11} + \frac23 I_{22} ,\\
    Q_{33} = - \frac13 I_{11} - \frac13 I_{22} ,\\
    Q_{21}=Q_{12} = I_{12},
\end{gather}
while all other $Q_{i j} $ are equal to zero. Now, from Newtonian mechanics $I_{ij}$ can be written as 
\begin{align}
    \ddot{\tilde{I}}_{i  j} (M,x;f_r)=
    4\left[\frac{x}{(1+x)^3} + \frac{1/x}{(1+1/x)^3}\right ]
    \notag\\
    \times\left(\frac{M}{2 \times 10^{12} M_{\odot}} \right  )^2
    \ddot{\tilde{I}}^{G}_{ij} (f_r),
\end{align}
where $x$ is the mass ratio of the two masses, and $I^{G}_{ij}$ is defined as $I_{ij}(M = 2 \times 10^{12} M_{\odot}, x=1 )$. Therefore, from the calculation of Section \ref{sec: single case}, we can extract the values of $\ddot{\tilde{I}}^{G}_{ij} (f_r)$  as
\begin{gather}
    \ddot{\tilde{I}}^{G}_{11} (f_r) = 2.86 \times 10^{21} 
    \,\left(\frac{ \text{Hz}}{f_r}\right)^2 \, kg\, m^2 s^{-1} ,\\
    \ddot{\tilde{I}}^{G}_{22} (f_r) =5.72 \times 10^{20} \, 
    \left(\frac{ \text{Hz}}{f_r}\right)^2 \, kg\, m^2 s^{-1} ,\\
    \ddot{\tilde{I}}^{G}_{12} (f_r) = \ddot{\tilde{I}}^{G}_{21} (f_r)  = 
    1.29 \times 10^{37} \, \left(\frac{ \text{Hz}}{f_r}\right) \, kg\, m^2 s^{-1} .
\end{gather}
Hence, inserting the above into \eqref{singleenergy} gives us the energy density of a single GW event.

The energy density can be written as follows
\begin{align}
\frac{d E(\xi)}{d \ln (f_r)} \propto M^{2} (C_1 f_{r} + C_2 \frac{1}{f_{r}}),
\end{align}
where $C_1, C_2$ are constants. When $f_{r} \geq 10^{-16}Hz$, the contribution of $\frac{1}{f_{r}}$ to the energy spectrum can be ignored. In this case, the energy spectrum is consistent with the result in \cite{2010gfe..book.....M} which is given by:
\begin{align}
\frac{d E(\xi)}{d \ln (f_r)} \propto f_{r} M^{2}.
\end{align}
It should be noted that integrating the energy spectrum of gravitational waves over the entire frequency range to determine the total energy released throughout the process will result in divergence.\\
The physical process of releasing high-frequency gravitational waves corresponds to two particles being very close together, causing rapid changes in the motion of particles. At this point, the release of gravitational waves will in turn have a significant impact on the motion of the two particles, rendering the approximation of elastic collision ineffective. Further more, it would even be unreasonable to consider the collision of dark matter halos as point particles under such circumstances.
Therefore, the frequency of the energy spectrum should be truncated at $f_{max} \approx v/b$ in order to avoid non-physical results. For $v \approx 300 km/s$ and $b \approx 10^{4}ly$, the cutoff frequency $f_{max} \approx 3.2 \times 10^{-15} Hz$.

 \subsection{Number density of GW sources} \label{subsec:number}
 
 Let us now calculate the number density of GW sources (per redshift, total mass and mass ratio interval) $\frac{d n}{d z dM dx}$. This number density is equal to the DM matter halos mergers rate, which can be calculated by combining the extended Press-Schechter (EPS) theory \cite{2010gfe..book.....M} and numerical simulations \cite{Genel:2010pb}:
\begin{align}
    \frac{d n}{d z dM dx} = n_{halo}(M,z)  \frac{d \omega}{d 
    z}\left(\frac{1}{n_{halo} } \frac{d n_{merger}}{d \omega  dx}  \right),
    \label{Numberdensity}
\end{align}
where $ n_{halo}(M,z)  $ is the number density  of dark matter halos  (per redshift per mass interval  in the co-moving space), $\omega = \omega(z)$ is a redshift-dependent function given below, and  $(\frac{1}{n_{halo} } \frac{d n_{merger}}{d \omega  dx}  )$ is the merger rate (at some $\omega$) for a pair of DM halos with fixed total mass $M$ and mass ratio $x$. In the following we handle these terms separately. 

We start with the definition of $\omega(z)$ \cite{2010gfe..book.....M}
\begin{align}
    \omega(z) = \frac{1.69}{D(z)},
    \label{omegazz}
\end{align}
where $D(z)$ is the linear growth rate of  matter density. $D(z)$ can be written as 
\begin{align}
    D(z) = \frac{1}{g(z=0)}\left[ \frac{g(z)}{1+z}\right] \label{Dz},
\end{align}
where a good approximation of $g(z)$ is
\begin{align}
    &g(z) \approx \frac{5}{2} \Omega_{\mathrm{m}}(z)
    \notag\\
    &\times\left\{\Omega_{\mathrm{m}}^{4 / 7}(z)-\Omega_{\Lambda}(z)+\left[1+\Omega_{\mathrm{m}}(z) /2\right]\left[1+\Omega_{\Lambda}(z) / 70\right]\right\}^{-1},
\end{align}
with $\Omega_{\Lambda}(z)$, $\Omega_{\mathrm{m}}(z)$ the density parameters of dark energy and matter sectors given by
\begin{align}
    \Omega_{\Lambda}(z)=\frac{\Omega_{\Lambda, 0}}{E^{2}(z)} ; \quad \Omega_{\mathrm{m}}(z)=\frac{\Omega_{\mathrm{m}, 0}(1+z)^{3}}{E^{2}(z)},
\end{align}
where the normalized Hubble function $E(z)\equiv H(z)/H_0$  reads as
\begin{align}
    E(z)\approx \left[\Omega_{\Lambda, 0}+\Omega_{\mathrm{m}, 0}(1+z)^{3}\right]^{1 / 2}, \label{Ez}
\end{align}
with the value of the Hubble function at present time given as \cite{Planck:2018nkj}
\begin{align}
    H_0  \approx 67.3 \, km\, s^{-1} Mpc^{-1}, \label{hubble}
\end{align}
and with the values $ \Omega_{\Lambda, 0},\Omega_{\mathrm{m}, 0}$ at present time taken as  \cite{Planck:2018nkj} 
\begin{align}
    \Omega_{\Lambda, 0} \approx 0.685,\\
    \Omega_{\mathrm{m}, 0} \approx 0.317.\label{OmegaMO}
\end{align}
Note that in the above we consider that the underlying cosmology is $\Lambda$CDM concordance scenario, i.e., the dark energy sector is the cosmological constant.

We continue by using the EPS theory in order to  write the formula of the number density of DM halos $n_{halo}$. We consider that the halos merge when the redshift is between $z$ and $z+dz$ , and that the emitted GW signals are detected at Earth at present. In co-moving space those halos are in the volume $\Delta V = 4 \pi r^2(z) d(r(z)) $. Now, the EPS theory provides the number density of DM halos $n_{EPS}(M,z)$ at some redshift $z$ and mass $M$. Therefore, we have 
\begin{align}
    n_{halo} = 4 \pi r^2(z) \frac{d r(z)}{d z} n_{EPS}(M,z),
    \label{nhaloeq}
\end{align} 
where the radius in the co-moving space $r(z)$ is  \cite{2010gfe..book.....M}
\begin{align}
    r(z) = \frac{c}{H_0} \int^{z}_{0} d z^{'} \frac{1}{E(z^{'})},
\end{align}
while the  formula of $n_{EPS}(M,z)$ is \cite{2010gfe..book.....M}
\begin{align}
     n_{EPS}(M,z) = 
    \sqrt{\frac{2}{\pi}} \frac{\bar{\rho}}{M^{2}} \frac{\delta_{\mathrm{c}}}{\sigma} 
    \exp \left(-\frac{\delta_{\mathrm{c}}^{2}}{2 
    \sigma^{2}}\right)\left|\frac{\mathrm{d} \ln \sigma}{\mathrm{d} \ln M}\right|. 
\end{align}
In the above expression $\bar{\rho} = \rho_c \Omega_{m,0}$ is the mean density of the matter component, $\delta_{\mathrm{c}} =\omega  =  \frac{1.69}{D(z)}$, while $\sigma(M)$ is the variance of the matter density perturbation  which can be estimated as \cite{2010gfe..book.....M}
\begin{align}
    \sigma(M)  \approx \sigma_8 \left (\frac{R}{r_8}\right)^{-\beta},
\end{align}
with $M = \frac{4\pi}{3} \bar{\rho} R^3$ , $\sigma_8 \approx 1$, $\beta \approx 0.6 + 0.8  (\Omega_{m,0}  h )$, $h = 0.673$ , and $r_8 = 8\,\text{Mpc}\, h^{-1}$, leading to  
\begin{align}
    \left|\frac{\mathrm{d} \ln \sigma}{\mathrm{d} \ln M}\right|=  \frac{\beta}{3}.
\end{align}

Finally, the last term of \eqref{Numberdensity}, namely  $(\frac{1}{n_{halo} } \frac{d n_{merger}}{d \omega  dx}  )$ (dimensionless since  both $\omega,x$ are  dimensionless), can be found in \cite{Genel:2010pb} and it is given by
\begin{align}
    \left(\frac{1}{n_{halo} } \frac{d n_{merger}}{d \omega  dx} \right ) = A 
    \left(\frac{M}{10^{12} M_{\odot}}  \right)^{\alpha} x^{b} \exp \left[(\tilde{x} 
    / x)^{\gamma}\right],
    \label{dnhalodn}
\end{align}
where the  best-fit parameters from   simulations are   $A=0.065$, $\alpha=0.15$, $b=-0.3$, $\tilde{x}=2.5$, $\gamma=0.5$ \cite{Genel:2010pb}.

In summary, inserting \eqref{omegazz}, \eqref{nhaloeq} and \eqref{dnhalodn} into \eqref{Numberdensity}, provides  the value of the number density of GW sources $\frac{dn}{dzdMdx}$.

\subsection{The energy spectrum of the stochastic gravitational wave background}

\begin{figure}[ht]
    \centering
    \includegraphics[width=1\columnwidth]{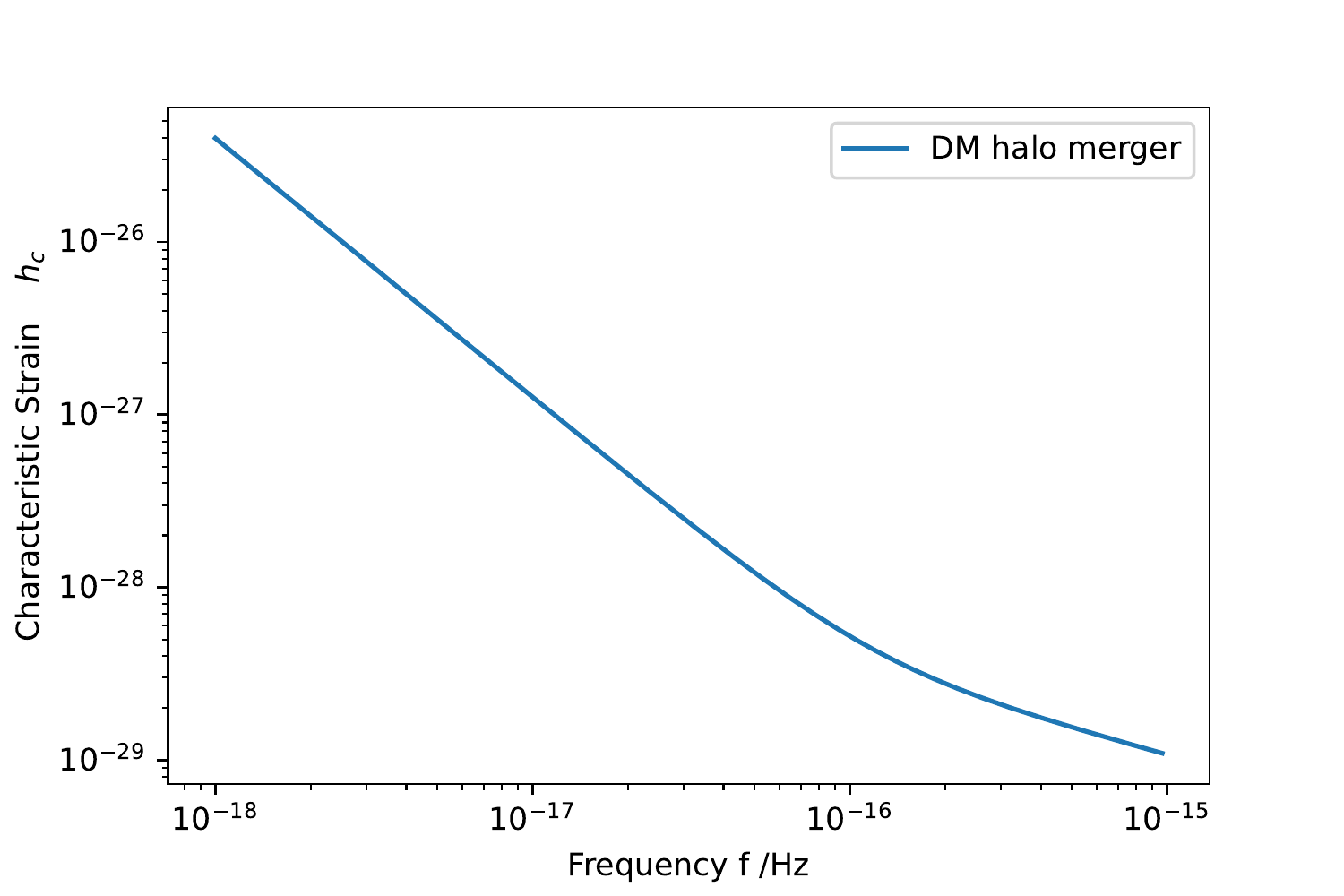}
    \caption{\it{The characteristic strain $h_{c}(f)$ as a function of the frequency  of the stochastic gravitational wave background created by DM halos, namely galaxies and galaxy clusters, collisions . }}
    \label{fig: GWB contribution}
\end{figure}

We have now all the ingredients needed in order to calculate the energy spectrum of the stochastic gravitational wave background. This is given by \eqref{finaleqshort}, in which the energy spectrum of a single GW event $\frac{d E(\xi)_{\mathrm{gw}}}{d \ln f_{r}}$ was calculated in subsection \ref{subsec:energy}, while the number density of GW sources $\frac{d n}{d z dM dx}$ was calculated in subsection \ref{subsec:number}. Assembling everything, we finally obtain the stochastic gravitational wave background resulting from  DM halos collisions in the universe, which is calculated numerically and it is shown in Fig. \ref{fig: GWB contribution}. Besdies, instantaneous collision approximation introduce a cutoff frequency, the frequency beyond which the signal is truncated is the inverse of the timescale of the collision. The cutoff frequency for instantaneous collisions is $f_{max} \approx  v/b$. For the case of the dark matter halos collisions, This frequency is about $3 \times 10^{-15} Hz$.

As we can see, the contribution of GW radiated from  the collisions of DM halos, namely galaxies and galaxy clusters,  is quite small comparing to other sources. In the  pulsar timing array (PTA) band, where $f \approx 10^{-9} Hz$, and where the current observational limit is  $h_{c} \approx 10^{-15}$ \cite{Burke-Spolaor:2018bvk}. we obtain an effect of the order of  $h_{c} \approx 10^{-29}$ in the band of $f \approx 10^{-15} Hz$. Nevertheless, in very low frequency band $h_{c}$ will be larger.  In general, with current observational sensitivity the effect of the DM halos collisions on the stochastic gravitational wave background cannot be detected\cite{Tinto:2001ii,Caprini:2019pxz,Karnesis:2022vdp,Chapman-Bird:2022tvu}. Note that one could try to extend the analysis, by considering, instead of point masses, a group of mass points with  Navarro, Frenk \& White (NFW) density profile \cite{2010gfe..book.....M} to simulate  DM halo collisions, nevertheless the results are expected to be at the same order of magnitude.

Dark matter halos are in reality extended objects and not point-particles. Strictly speaking, this two point toy model only suits for the beginning of the merger of 2 DM halos, at this stage the mechanical energy of the 2 mass centers of DM halo is approximately a constant. One may use N-body simulation to calculate the GW radiated from the merger more precisely. However, as an estimation of order of magnitude of the contribution to the GWB, the simple model in this paper is good enough.

\section{Conclusions}
\label{sec: conclusion}

In this work we investigated the effect of the dark matter halos collisions, namely collisions of galaxies and galaxy clusters, through gravitational bremsstrahlung, on the  stochastic gravitational wave background.

In order to achieve this goal, we first calculated the gravitational wave signal of a single DM halo collision event. As an estimation of the order of magnitude, we handled the two DM halos as mass points. Furthermore,  since the strength of such GW signals is weak, we adopted linear perturbation theory of General Relativity, namely we extracted the GW signal using the second time derivative of the quadruple moment. Additionally, since the velocity of DM halos is small, we applied non-relativistic Newtonian Mechanics. Hence, we extracted the GW signal through bremsstrahlung from a single  DM halo collision. As we showed, $\bar{h}_{ij}$ is of the order of $10^{-22}$, and it becomes maximum at the time of shortest distance as expected. However, since such an event typically  corresponds to duration of the order of $10^{15} s$, we deduce that a single signal of this kind of GW is extremely hard to be detected.

As a next step we proceeded to the  calculation of the energy spectrum of the collective effect of all DM halos collisions in the Universe. This can arise by the energy spectrum of a GW signal radiated by a single collision, multiplied by the DM halo collision rate, and integrating over the whole Universe. Firstly, knowing the signal of a single collision we calculated its energy spectrum. Secondly, concerning the DM halo collision rate we showed that it is given by the product of the number density of DM halos, which is calculated by the EPS theory, with  the collision rate of a single DM halo, which is given by simulation results, with a function of the linear growth rate of matter density through cosmological evolution. Hence, integrating over all mass and distance ranges, we finally extracted the spectrum of the stochastic gravitational wave background created by DM halos collisions.

As we show, the resulting  contribution to the stochastic gravitational wave background is of the order of $h_{c} \approx 10^{-29}$ in the band of $f \approx 10^{-15} Hz$. However, in very low frequency band, $h_{c}$ is larger.  With current observational sensitivity it cannot be detected.

In summary, with the current and future significant advance in gravitational-wave astronomy, and in particular with the tremendous improvement on the sensitivity bounds that Collaborations like Laser Interferometer Space Antenna (LISA), Einstein Telescope (ET), Cosmic Explorer (CE), etc will bring, it is both interesting and necessary to investigate all possible contributions to the stochastic gravitational wave background. And the gravitational bremsstrahlung during  galaxies and galaxy clusters collisions is one of them.

\section*{Acknowledgments}

We are grateful to Yifu Cai, Jiewen Chen, Zihan Zhou, Jiarui Li and Yumin Hu, Bo Wang and Rui Niu for helpful discussions. This work is supported in part by the National  Key R\&D Program of China (2021YFC2203100), by CAS young interdisciplinary 
innovation team (JCTD-2022-20), by the NSFC (12261131497), by 111 Project for ``Observational and 
Theoretical Research on Dark Matter and Dark Energy'' (B23042), by the Fundamental Research Funds for Central Universities, by the CSC Innovation Talent Funds, by the USTC Fellowship for International Cooperation, and by the USTC Research Funds of the Double First-Class Initiative. ENS acknowledges participation in the COST Association Action CA18108 ``{\it Quantum Gravity Phenomenology in the Multimessenger Approach (QG-MM)}''. All numerics were operated on the computer clusters {\it LINDA} \& {\it JUDY} in the particle cosmology group at USTC.

\bibliographystyle{spphys}
\bibliography{reference}

\end{document}